\documentclass[twocolumn,aps,prl,superscriptaddress,showpacs]{revtex4-1}
\usepackage{graphicx}
\usepackage{epsfig}
\usepackage{amsmath}
\usepackage{amssymb}
\usepackage{latexsym}
\usepackage{array}

\def \a{\alpha}

\def \be{\begin{equation}}
\def \ee{\end{equation}}
\def \ben{\begin{eqnarray}}
\def \een{\end{eqnarray}}

\begin{document}

\title{Cosmology in presence of dark energy in an emergent gravity scenario}

\author{Debashis Gangopadhyay}
\altaffiliation{debashis@rkmvu.ac.in}
\affiliation{Department of Physics, Ramakrishna Mission Vivekananda University, Belur Matth, Howrah}

\author{Goutam Manna}
\altaffiliation{goutammanna.pkc@gmail.com}
\affiliation{Department of Physics, Prabhat Kumar College, Contai, Purba Medinipur-721401, West Bengal, India}

\date{\today}

\begin{abstract}   
We obtain the analogues of the Friedman equations in an emergent gravity scenario in the presence of dark energy.
The background metric is taken to be Friedman-Lemaitre-Robertson-Walker (FLRW). We show that
if $\dot\phi ^{2}$ is the dark energy density (in units of the critical density) then 
(a) for total energy density greater than the pressure  (non-relativistic scenario, matter domination)
the deceleration parameter $q(t)\approx\frac {1}{2} [1 + 27 \dot\phi ^{2}+...] > \frac{1}{2}$ 
(b) for total energy density equal to 3 times the pressure (relativistic case, radiation 
domination), the deceleration parameter $q(t)\approx 1 + 18\dot\phi ^{2} +... >  1$  and
(c) for total energy density equal to the negative of the pressure (dark energy scenario), the deceleration parameter $q(t)< -1$. Our results indicate that many aspects 
of standard cosmology can be accommodated with the presence of dark energy  right 
from the beginning of the universe where the time parameter $t\equiv \frac{t}{t_{0}}$, $t_{0}$
being the present epoch. 

\end{abstract}


\maketitle

\section{Introduction}

Scalar fields $\phi$ whose lagrangians are non-canonical are candidates for $k-$essence fields that give rise to dark energy. The general form for such  lagrangians is proportional to $F(X)$ with $X={1\over 2}g^{\mu\nu}\nabla_{\mu}\phi\nabla_{\nu}\phi$.
Relevant literature for such fields in  cosmology, inflation, dark matter, dark energy 
and strings can be found in \cite{gor1}-\cite{gibbons}.

A natural question is whether the standard cosmology is modified if we take into account the presence of dark energy while building up the Friedman equations. In this context it has to be further remembered that dynamical solutions of the $k$-essence equation of motion changes the metric for the perturbations around these solutions \cite{wald}-\cite{babi3}. The perturbations propagate in an emergent spacetime with metric $\tilde G^{\mu\nu}$ different from (and also not conformally equivalent to) the gravitational metric $g^{\mu\nu}$. $\tilde G^{\mu\nu}$ now depends on $\phi$. Hence we have to determine the analogue of the Friedman equations in the context of dark energy in an emergent gravity scenario. 

The motivation of this work is to seek plausible answers to the above question. 
Taking the background metric to be FLRW, 
we obtain the modifications of the standard cosmological parameters in  the radiation dominated, matter dominated and dark energy dominated phases of the universe. The dark energy density is identified with the kinetic energy $\dot\phi ^{2}$ of the $k-$essence field. The  standard cosmological  parameters are retrieved when $\dot\phi^{2}\rightarrow 0$, i.e., the dark energy vanishes.

The plan of the paper is as follows: section 2 introduces emergent gravity concepts, section 3 discusses the emergent gravity equations of motion, section 4 contains the analogues of the Friedman equations in the presence of dark energy, section 5 discusses the solutions of the resulting cosmological equations arising from emergent gravity equations of motion and the analogue of the Friedman equations and section 6 comprises of our conclusions. 
\section{Emergent Gravity}
The $k-$essence scalar field $\phi$ is minimallly coupled to the gravitational 
field $g_{\mu\nu}$ and  the action is:  
\ben
S_{k}[\phi,g_{\mu\nu}]= \int d^{4}x {\sqrt -g} L(X,\phi)
\label{eq:1}
\een
where $X={1\over 2}g^{\mu\nu}\nabla_{\mu}\phi\nabla_{\nu}\phi$ 
and the energy momentum tensor for the $k-$essence scalar field is:
\ben
T_{\mu\nu}\equiv {2\over \sqrt {-g}}{\delta S_{k}\over \delta g^{\mu\nu}}
= L_{X}\nabla_{\mu}\phi\nabla_{\nu}\phi - g_{\mu\nu}L
\label{eq:2}
\een
$L_{\mathrm X}= {dL\over dX},~~ L_{\mathrm XX}= {d^{2}L\over dX^{2}},
~~L_{\mathrm\phi}={dL\over d\phi}$ and  
$\nabla_{\mu}$ is the covariant derivative defined with respect to the metric $g_{\mu\nu}$.
The equation of motion for the scalar field is \cite{vikman}:
\ben
-{1\over \sqrt {-g}}{\delta S_{k}\over \delta \phi}= \tilde G^{\mu\nu}\nabla_{\mu}\nabla_{\nu}\phi +2XL_{X\phi}-L_{\phi}=0
\label{eq:3}
\een

where the effective metric $\tilde G^{\mu\nu}$ is 
\ben
\tilde G^{\mu\nu}(\phi,\nabla \phi)\equiv L_{X} 
 g^{\mu\nu} + L_{XX} \nabla ^{\mu}\phi\nabla^{\nu}\phi
\label{eq:4}
\een
and is physically meaningful only when $1+ {2X  L_{XX}\over L_{X}} > 0$.

We first carry out the conformal transformation
$G^{\mu\nu}\equiv {c_{s}\over L_{x}^{2}}\tilde G^{\mu\nu}$, with
$c_s^{2}(X,\phi)\equiv{(1+2X{L_{XX}\over L_{X}})^{-1}}\equiv sound ~ speed $.
Then  the inverse metric of $G^{\mu\nu}$ is  
\ben G_{\mu\nu}={L_{X}\over c_{s}}[g_{\mu\nu}-{c_{s}^{2}}{L_{XX}\over L_{X}}\nabla_{\mu}\phi\nabla_{\nu}\phi] 
\label{eq:5}
\een
A further conformal transformation $\bar G_{\mu\nu}\equiv {c_{s}\over L_{X}}G_{\mu\nu}$ gives
\ben \bar G_{\mu\nu}
={g_{\mu\nu}-{{L_{XX}}\over {L_{X}+2XL_{XX}}}\nabla_{\mu}\phi\nabla_{\nu}\phi}
\label{eq:6}
\een
$L_{X}\neq 0$ always for the sound speed 
$c_{s}^{2}$ to be positive definite and only 
then equations $(1)-(4)$ is physically meaningful. 
This can be seen as follows.   $L_{X}=0$ implies that $L$ does not 
depend on $X$ so that in equation (\ref{eq:1}),  $L(X,\phi)\equiv L(\phi)$.
So the $k-$essence lagrangian $L$ becomes pure potential.
Then the very definition of $k-$essence fields is meaningless because 
such fields correspond to lagrangians where the kinetic energy dominates 
over the potential energy. So lagrangian cannot be pure potential.
Also, the very concept of minimally coupling the $k-$essence field $\phi$ to 
the gravitational field $g_{\mu\nu}$ becomes redundant and 
equation (\ref{eq:1}) meaningless and equations (\ref{eq:4}-\ref{eq:6}) ambiguous.

Non-trivial configurations imply 
$\partial_{\mu}\phi\neq 0$ and  $\bar G_{\mu\nu}$ not conformally 
equivalent to $g_{\mu\nu}$. So this $k-$essence
field has properties different from canonical 
scalar fields defined with $g_{\mu\nu}$ and the local causal 
structure is also different from those defined with $g_{\mu\nu}$.
Further, if $L$ is not an explicit function of $\phi$
then equation $(3)$ becomes ;
\ben
-{1\over \sqrt {-g}}{\delta S_{k}\over \delta \phi}
= \bar G^{\mu\nu}\nabla_{\mu}\nabla_{\nu}\phi=0
\label{eq:7}
\een

We take $L=L(X)=1-V\sqrt{1-2X}$ \cite{born}. 
This is a particular case of the Born-Infeld lagrangian 
$L(X,\phi)= 1-V(\phi)\sqrt{1-2X}$ for $V(\phi)=V=constant$
and  $V<<kinetic ~ energy ~ of~\phi$ i.e.$V<< (\dot\phi)^{2}$. 
Then $c_{s}^{2}(X,\phi)=1-2X$.
For scalar fields $\nabla_{\mu}\phi=\partial_{\mu}\phi$. 
So (\ref{eq:6}) becomes
\ben
\bar G_{\mu\nu}= g_{\mu\nu} - \partial _{\mu}\phi\partial_{\nu}\phi
\label{eq:8}
\een
Note that the first conformal transformation is used to 
identify the inverse metric $G_{\mu\nu}$.
The second conformal transformation realises the mapping onto the   
metric given in $(8)$ for $L(X)=1 -V\sqrt{1-2X}$.

Consider the second conformal transformation $\bar G_{\mu\nu}\equiv {c_{s}\over L_{X}}G_{\mu\nu}$.
Following \cite{wald} the new Christoffel symbols are related to the old ones by  
\ben
\bar\Gamma ^{\alpha}_{\mu\nu} 
=\Gamma ^{\alpha}_{\mu\nu} + (1-2X)^{-1/2}G^{\alpha\gamma}[G_{\mu\gamma}\partial_{\nu}(1-2X)^{1/2}
\nonumber\\
+G_{\nu\gamma}\partial_{\mu}(1-2X)^{1/2}-G_{\mu\nu}\partial_{\gamma}(1-2X)^{1/2}]
\nonumber\\
=\Gamma ^{\alpha}_{\mu\nu} -\frac {1}{2(1-2X)}[\delta^{\alpha}_{\mu}\partial_{\nu}X
+ \delta^{\alpha}_{\nu}\partial_{\mu}X]
\label{eq:9}
\een
The second term on the right hand side is symmetric under exchange of $\mu$ and $\nu$ 
so that the symmetry of $\bar\Gamma$ is maintained. The second term originates from the 
$k-$essence lagrangian and this additional term signifies additional interactions (forces). 
The new geodesic equation in terms of $\bar\Gamma$ now is 
\ben
\frac {d^{2}x^{\alpha}}{d\tau^{2}} +  \bar\Gamma ^{\alpha}_{\mu\nu}\frac {dx^{\mu}}{d\tau}\frac {dx^{\nu}}{d\tau}=0
\label{eq:10}
\een

\section{Emergent equations of motion for FLRW gravitational metric}
Take the gravitational metric $g_{\mu\nu}$ to be FLRW and assume
that the $k-$essence scalar field $\phi(r,t)$ is spherically symmetric
($\partial_{t}\phi=\partial_{0}\phi=\dot\phi$ and $\partial_{r}\phi=\partial_{1}\phi=\phi'$).
Then (8) becomes

\ben
\bar G_{00}=g_{00}-(\partial _{0}\phi)^{2}=1-\dot \phi^2\nonumber\\
\bar G_{11}=g_{11}-(\partial _{r}\phi)^{2}=-{a^{2}(t)\over{1-Kr^{2}}}-(\phi ') ^{2}\nonumber\\
\bar G_{22}=g_{22}=-a^{2}(t)r^{2}\nonumber\\
\bar G_{33}=g_{33}=-a^{2}(t)r^{2}sin^{2}\theta\nonumber\\
\bar G_{01}=\bar G_{10}=-\dot\phi\phi '
\label{eq:11}
\een
where the FLRW metric components are
$ g_{00}=1; g_{11}=-{a^{2}(t)\over{1-Kr^{2}}}; g_{22}=-a^{2}(t)r^{2}; 
g_{33}=-a^{2}(t)r^{2}sin^{2}\theta; g_{ij} (i\neq j)=0$.

The line element becomes
\ben
ds^{2}=(1-\dot \phi^2)dt^{2}-({a^{2}\over{1-Kr^{2}}}+(\phi ') ^{2})dr^{2}\nonumber\\
-2\dot\phi\phi 'dtdr-a^{2}r^{2}d\Omega^{2}\label{eq:12}
\een
with $d\Omega^{2}=d\theta^{2}+sin^{2}\theta d\phi^{2}$.

Consider a co-ordinate transformation from $(t,r,\theta,\phi)$ to 
$(\omega,r,\theta,\phi)$ so that \cite{wein1}:
\ben
d\omega=dt-({{\dot\phi\phi '}\over{1- \dot\phi ^{2}}})dr
\label{eq:13}
\een

Then (\ref{eq:12}) becomes
\ben
ds^{2}=(1 - \dot\phi ^{2})d\omega^{2}
-[{a^{2}\over{1-Kr^{2}}}+(\phi ') ^{2}\nonumber\\+{{(\dot\phi\phi ')^{2}}\over{(1-\dot\phi^{2})}}]dr^{2}
-a^{2}r^{2}d\Omega^{2}\nonumber\\
\label{eq:14}.
\een
i.e.
\ben
\fontsize{6pt}{8pt}
\bar G_{\mu\nu} = \left(\begin{array}{cccc}
{(1 - \dot\phi ^{2})} & 0 & 0 & 0\\
0 & -{Z} & 0 & 0\\
0 & 0 & {-(a^{2}r^{2})} & 0\\
0 & 0 & 0 & {-(a^{2}r^{2}sin^{2}\theta)} \\
\end{array}\right)
\label{eq:15}
\een
while its inverse is 
\ben
\fontsize{6pt}{8pt}
\bar G^{\mu\nu} = \left(\begin{array}{cccc}
{(1 - \dot\phi ^{2})^{-1}} & 0 & 0 & 0\\k
0 & -{Z^{-1}} & 0 & 0\\
0 & 0 & {-(a^{2}r^{2})^{-1}} & 0\\
0 & 0 & 0 & {-(a^{2}r^{2}sin^{2}\theta)^{-1}} \\
\end{array}\right)\nonumber\\
\label{eq:16}
\een
with $Z={({a^{2}\over{1-Kr^{2}}}+(\phi ') ^{2}+{{(\dot\phi\phi ')^{2}}\over{(1-\dot\phi^{2})}})}$.

The equation (\ref{eq:7}) means
\ben
\bar G^{00}\partial^{2}_{0}\phi+\bar G^{11}(\partial^{2}_{1}\phi-{a\dot a\over{1-Kr^{2}}}\partial_{0}\phi
\nonumber\\-{Kr\over{1-Kr^{2}}}\partial_{1}\phi)=0
\label{eq:17}
\een
i.e.
\ben
\ddot\phi[a^{2}(1-\dot\phi^{2})+(\phi')^{2}(1-Kr^{2})]\nonumber\\
=(1-\dot\phi^{2})^{2}[\phi''(1-Kr^{2})-a\dot a \dot\phi-Kr\phi']
\label{eq:18}
\een
We shall, henceforth, consider the FLRW universe for homogeneous dark energy fields only. So   
\ben
\phi(r,t)\equiv \phi(t)
\label{eq:19}
\een
Here $\dot\phi^{2}\neq 0$ since the  $k-$essence field must have  non-zero kinetic energy.
Also $\dot\phi^{2}\neq 1$ because $\Omega_{matter} +\Omega_{radiation} +\Omega_{dark energy}= 1$ 
and $\dot\phi^{2}$  measured in units of the critical density is nothing but $\Omega_{dark energy}$.
Further, $\dot\phi ^{2}<1$ always in order that the  signature 
of the metric (\ref{eq:15}) does not become ill-defined.  
Therefore $0<\dot\phi ^{2}<1$.
Therefore (\ref{eq:18}) becomes
\ben
{\dot a\over{a}}=H(t)=-{\ddot\phi\over{\dot\phi(1-\dot\phi ^{2})}}
\label{eq:20}
\een
where $H(t)={\dot a\over{a}}$ is Hubble parameter ({\it always $\dot a\neq 0$}).
So the equations of motion of emergent gravity relate the Hubble parameter to 
time derivatives of the $k-$ essence scalar field.
 
\section{The analogue of Friedmann Equations in presence of Dark Energy} 

Using metrics (\ref{eq:15} and \ref{eq:16})for homogeneous fields $\phi (t)$
we get the non-vanishing connection coefficients as:

$$\bar \Gamma_{00}^{0}=-{\dot\phi~\ddot\phi\over{1-\dot\phi ^{2}}};\nonumber\\$$
$$\bar \Gamma_{11}^{0}={1\over{1-\dot\phi ^{2}}}{a\dot a\over{1-Kr^{2}}};\nonumber\\$$
$$\bar \Gamma_{22}^{0}={a \dot a r^{2}\over{1-\dot\phi ^{2}}};\nonumber\\$$
$$\bar \Gamma_{33}^{0}={a \dot a r^{2}sin^{2}\theta\over{1-\dot\phi ^{2}}};\nonumber\\$$
$$\bar \Gamma_{01}^{1}=\bar \Gamma_{10}^{1}={\dot a\over{a}};\nonumber\\$$
$$\bar \Gamma_{11}^{1}={Kr\over{1-Kr^{2}}};\nonumber\\$$
$$\bar \Gamma_{22}^{1}=-r(1-Kr^{2});\nonumber\\$$
$$\bar \Gamma_{33}^{1}=-r sin^{2}\theta(1-Kr^{2});\nonumber\\$$
$$\bar \Gamma_{02}^{2}=\bar \Gamma_{20}^{2}={\dot a\over{a}};\nonumber\\$$
$$\bar \Gamma_{12}^{2}=\bar \Gamma_{21}^{2}={1\over{r}};\nonumber\\$$
$$\bar \Gamma_{33}^{2}=-sin\theta~cos\theta;\nonumber\\$$
$$\bar \Gamma_{03}^{3}=\bar \Gamma_{30}^{3}={\dot a\over{a}};\nonumber\\$$
$$\bar \Gamma_{13}^{3}=\bar \Gamma_{31}^{3}={1\over{r}};\nonumber\\$$
$$\bar \Gamma_{23}^{3}=\bar \Gamma_{32}^{3}=cot\theta.$$
Now we calculate the diagonal components of Ricci tensor for homogeneous scalar field since off-diagonal
components of Ricci tensor are zero.
\ben
\bar R_{00}=3{\ddot a\over{a}}+3{\dot a\over{a}}{\dot\phi\ddot\phi\over{(1-\dot\phi^{2})}}
\label{eq:21}
\een
\ben
\bar R_{11}=-{a^{2}\over{1-Kr^{2}}}[{\ddot a\over{a}}{1\over{(1-\dot\phi^{2})}}
+2{\dot a^{2}\over{a^{2}}}{1\over{(1-\dot\phi^{2})}}\nonumber\\+2{K\over{a^{2}}}
+{\dot a\over{a}}{\dot\phi\ddot\phi\over{(1-\dot\phi^{2})^{2}}}]
\label{eq:22}
\een
\ben
\bar R_{22}=-a^{2}r^{2}[{\ddot a\over{a}}{1\over{(1-\dot\phi^{2})}}
+2{\dot a^{2}\over{a^{2}}}{1\over{(1-\dot\phi^{2})}}+2{K\over{a^{2}}}\nonumber\\
+{\dot a\over{a}}{\dot\phi\ddot\phi\over{(1-\dot\phi^{2})^{2}}}]\nonumber\\
\label{eq:23}
\een
\ben
\bar R_{33}=-a^{2}r^{2}sin^{2}\theta[{\ddot a\over{a}}{1\over{(1-\dot\phi^{2})}}
+2{\dot a^{2}\over{a^{2}}}{1\over{(1-\dot\phi^{2})}}\nonumber\\+2{K\over{a^{2}}}
+{\dot a\over{a}}{\dot\phi\ddot\phi\over{(1-\dot\phi^{2})^{2}}}]
\label{eq:24}
\een
To calculate Ricci scalar:
\ben
\bar R_{0}^{0}=\bar G^{00}\bar R_{00}={1\over{(1-\dot\phi^{2})}}[3{\ddot a\over{a}}
+3{\dot a\over{a}}{\dot\phi\ddot\phi\over{(1-\dot\phi^{2})}}]
\label{eq:25}
\een
\ben
\bar R_{1}^{1}=\bar G^{11}\bar R_{11}=[{\ddot a\over{a}}{1\over{(1-\dot\phi^{2})}}
+2{\dot a^{2}\over{a^{2}}}{1\over{(1-\dot\phi ^{2})}}\nonumber\\+2{K\over{a^{2}}}
+{\dot a\over{a}}{\dot\phi\ddot\phi\over{(1-\dot\phi ^{2})^{2}}}]\nonumber\\
\label{eq:26}
\een
\ben
\bar R_{2}^{2}=\bar G^{22}\bar R_{22}=[{\ddot a\over{a}}{1\over{(1-\dot\phi ^{2})}}
+2{\dot a^{2}\over{a^{2}}}{1\over{(1-\dot\phi ^{2})}}\nonumber\\+2{K\over{a^{2}}}
+{\dot a\over{a}}{\dot\phi\ddot\phi\over{(1-\dot\phi^{2})^{2}}}]\nonumber\\
\label{eq:27}
\een
\ben
\bar R_{3}^{3}=\bar G^{33}\bar R_{33}=[{\ddot a\over{a}}{1\over{(1-\dot\phi^{2})}}
+2{\dot a^{2}\over{a^{2}}}{1\over{(1-\dot\phi^{2})}}\nonumber\\+2{K\over{a^{2}}}
+{\dot a\over{a}}{\dot\phi\ddot\phi\over{(1-\dot\phi^{2})^{2}}}]\nonumber\\
\label{eq:28}
\een
Therefore the Ricci Scalar:
\ben
\bar R=\bar R_{0}^{0}+\bar R_{1}^{1}+\bar R_{2}^{2}+\bar R_{3}^{3}\nonumber\\
=6[{\ddot a\over{a}}{1\over{(1-\dot\phi^{2})}}
+{\dot a^{2}\over{a^{2}}}{1\over{(1-\dot\phi^{2})}}+{K\over{a^{2}}}\nonumber\\
+{\dot a\over{a}}{\dot\phi\ddot\phi\over{(1-\dot\phi^{2})^{2}}}]
\label{eq:29}
\een
We have the Einstein's Field Equation:
$\bar E_{\mu\nu}=\bar R_{\mu\nu}-{1\over{2}}\bar G_{\mu\nu}\bar R=-8\pi G T_{\mu\nu}$
i.e.
\ben
\bar E_{\mu}^{\nu}=\bar R_{\mu}^{\nu}-{1\over{2}}\delta_{\mu}^{\nu}\bar R=-8\pi GT_{\mu}^{\nu}
\label{eq:30}
\een
where $G$ is gravitational constant and $T_{\mu\nu}$ is energy-momentum tensor.
Components of Einstein tensor are:
\ben
\bar E_{0}^{0}=-3[{\dot a^{2}\over{a^{2}}}{1\over{(1-\dot\phi^{2})}}+{K\over{a^{2}}}]
\label{eq:31};
\een
\ben
\bar E_{1}^{1}=\bar E_{2}^{2}=\bar E_{3}^{3}=-[2{\ddot a\over{a}}{1\over{(1-\dot\phi^{2})}}
+{\dot a^{2}\over{a^{2}}}{1\over{(1-\dot\phi^{2})}}\nonumber\\+{K\over{a^{2}}}
+2{\dot a\over{a}}{{\dot\phi\ddot\phi}\over{(1-\dot\phi^{2})^{2}}}]\nonumber\\
\label{eq:32}
\een
The energy-momentum tensor of an ideal fluid is
\ben
T_{\mu}^{\nu}=(p+\rho)u_{\mu}u^{\nu}-\delta_{\mu}^{\nu}p
\label{eq:33}
\een
where $p$ is pressure and $\rho$ is the matter density of the cosmic fluid.
In the co-moving frame  we have $u^{0}=1$ and $u^{\alpha}=0~;~\alpha=1,2,3$.

Now the general $k-$essence field theoretic lagrangian $L(X,\phi)$, which explicitly 
depends on $\phi$, is not equivalent to isentropic hydrodynamics because $\phi$ and $X$ 
are independent and hence the pressure cannot be a function of the energy density $\rho$ 
only. So a pertinent question is whether we are at all justified in assuming a perfect fluid 
model when dark energy is present. The answer is yes because our lagrangian $L(X)=1-V\sqrt{1-2X}$,
where $V$ is a constant, does not depend explicitly on $\phi$. This class of models is 
equivalent to perfect fluid models with zero vorticity and the pressure (lagrangian) can be  
expressed through the energy density only \cite{vikman}.  

Then (\ref{eq:33}) becomes
\ben
T_{0}^{0}=\rho~;~T_{1}^{1}=T_{2}^{2}=T_{3}^{3}=-p
\label{eq:34}
\een
Using equations (\ref{eq:30})-(\ref{eq:33}) we get 
\ben
\rho_{d}={3\over{8\pi G}}[{\dot a^{2}\over{a^{2}}}{1\over{(1-\dot\phi^{2})}}+{K\over{a^{2}}}]
\label{eq:35}
\een
and
\ben
p_{d}=-{1\over{8\pi G}}[2{\ddot a\over{a}}{1\over{(1-\dot\phi^{2})}}
+{\dot a^{2}\over{a^{2}}}{1\over{(1-\dot\phi^{2})}}+{K\over{a^{2}}}\nonumber\\
+2{\dot a\over{a}}{\dot\phi\ddot\phi\over{(1-\dot\phi^{2})^{2}}}]\nonumber\\
\label{eq:36}
\een
where we now replace $\rho$ by $\rho_{d}$ as the total matter density in presence of dark energy and $p$ by $p_{d}$ as the pressure when dark energy  is present. Note that both 
$\rho_{d}, p_{d}$  reduce  to  the usual  quantities $\rho, p$ \cite{wein2,mukhanov} when  dark  energy is absent,  i.e.,  $(\dot\phi)^{2} = 0$. 
The usual Friedman equations are now modified into the above two equations (\ref{eq:35} and \ref{eq:36}) in the presence of $k-$essence scalar field $\phi$.

Combining above two equations (\ref{eq:35} and \ref{eq:36}) we get,
\ben
{4\pi G\over{3}}(\rho_{d}+3p_{d})
=-[{\ddot a\over{a}}{1\over{(1-\dot\phi^{2})}}+{\dot a\over{a}}{\dot\phi\ddot\phi_{2}\over{(1-\dot\phi^{2})^{2}}}]
\label{eq:37}
\een 
Now differentiating equation (\ref{eq:35}) with respect to cosmic time $t$ 
and substituting the result in equation (\ref{eq:37}) we get,
\ben
\dot\rho_{d}=-3{\dot a\over{a}}(p_{d}+\rho_{d})=-3H(p_{d}+\rho_{d})
\label{eq:38} 
\een
which is the required energy conservation equation in presence of dark energy.
Again it may be noted that one recovers the usual energy conservation equation 
\cite{wein2,mukhanov} when dark energy is absent.

Now assume  that the criterion for  non-relativistic scenario remains the same, 
{\it viz.}, $\rho_{d}\gg p_{d}$.
We restrict now to $K=0$ as observationally this is most likely. Then the 
above condition reduces to 
$\frac{\dot a^{2}}{a^{2}}\gg - \frac{\ddot a}{2a}-\frac{\dot a\dot\phi\ddot\phi}{2a(1-\dot\phi^{2})}$.
So the second term on right hand side must be always positive i.e. $\dot\phi\ddot\phi$ must be always negative. This means that $\frac{d\dot\phi^{2}}{dt} < 0$. This criterion is consistent with the fact 
that the dark energy density cannot increase in a matter dominated era. 

Then, neglecting $p_{d}$ in (\ref{eq:38}) gives 
$\dot\rho_{d}\frac{a}{\dot a}+3\rho_{d}=0$ which has the solution  
\ben
\rho_{d}^{mat}=\frac{A}{a^{3}}
\label{eq:39}
\een
where $A=constant$. Assuming that the total energy i.e. $\rho_{d} a^3$ is a constant,
we equate this to the present epoch energy i.e. $ \rho_{d} a^3=\rho_{d0} a_{0}^3$, where 
$\rho_{d0}$ and $a_{0}$ are the matter density and scale radius at the present epoch ($t=t_{0}$). This fixes the constant $A=\rho_{d0}a_{0}^{3}$ in terms of present epoch values.

For the relativistic situation we assume again that the criterion is same as in standard cosmology,
i.e. $p_{d}=\frac{\rho_{d}}{3}$. Here this gives the condition 
$\frac{\dot a^{2}}{a^{2}}= - \frac{\ddot a}{a}-\frac{\dot a\dot\phi\ddot\phi}{a(1-\dot\phi^{2})}$
For same reasons as given in the previous case, here also the conditions are consistent.

We get from (\ref{eq:38})the solution 
\ben
\rho_{d} ^{rad}=\frac{B}{a^{4}}
\label{eq:40}
\een
where the constant  $B$ is fixed to be $B=\rho_{d0}a_{0}^{4}$ following same arguments as before.

Finally we consider the dark energy dominated scenario $p_{d}= -\rho_{d}$,  i.e.,
$\frac{\dot a^{2}}{a^{2}} = \frac{\ddot a}{a} + \frac{\dot a\dot\phi\ddot\phi}{a(1-\dot\phi^{2})}$
This means that here $\frac {d\dot\phi^{2}}{dt} > 0 $ i.e. the dark energy density must increase. 
This is also consistent. Equation (\ref{eq:38}) then leads to 
\ben
\rho_{d}=W
\label{eq:41}
\een
where we may choose the constant $W$ to be  $\dot\phi ^{2} |_{t=t'}$ 
with $t'$ denoting some specific epoch.

Therefore, the difference from the standard cosmology lies only in the fact that in our case the 
dark energy density  (which is being identified with the kinetic energy of the $k-$essence field) 
has the following behaviour: \textit{in the matter and radiation dominated eras the time rate of change of dark energy density decreases, while in the dark energy dominated epoch this rate increases.}

\section{Solutions of the modified equations}
\subsection{Non-relativistic case (Matter dominated Universe)}
The non-relativistic case means $\rho_{d}\gg p_{d}$ and   
(\ref{eq:35}) and (\ref{eq:36}) can be written as follows,
\ben
{H^{2}\over{(1-\dot\phi ^{2})}}+{K\over{a^{2}}}={8\pi G\over{3}}{A\over{a^{3}}}
\label{eq:42}
\een
and 

\ben
2{\ddot a\over{a}}{1\over{(1-\dot\phi^{2})}}
+{H^{2}\over{(1-\dot\phi^{2})}}+{K\over{a^{2}}}
+2H{\dot\phi\ddot\phi\over{(1-\dot\phi^{2})^{2}}}=0\nonumber\\
\label{eq:43}
\een
Eliminating $A$ using (\ref{eq:39}), and remembering that for $t=t_{0}$ (present epoch) $\rho_{d}=\rho_{d0}, a=a_{0}, H=H_{0}$  (\ref{eq:42}) becomes
\ben
{K\over{a_{0}^{2}}}={8\pi G\over{3}}[\rho_{d0}-\rho_{d}^{c}]
\label{eq:44}
\een
where 
\ben
\rho_{d}^{c}={3H_{0}^{2}\over{8\pi G(1-\dot\phi^{2})}}
\label{eq:45} 
\een
{\it is the critical value of matter density when dark energy is present.}
For  $\dot\phi^{2} < 1$   (\ref{eq:45}) becomes
\ben
\rho_{d}^{c}=\rho_{c}+\rho_{c}\dot\phi^{2}
\label{eq:46}
\een
keeping terms upto $O(\dot\phi ^2)$ only. Here $\rho_{c}={3H_{0}^{2}\over{8\pi G}}$, the  critical value of matter density and $\rho_{d}^{c}>\rho_{c}$.

Now  consider the FLRW universe with $K=0$. We get from (\ref{eq:44}) 
\ben
\rho_{d0} = \rho_{d}^{c} = {{3H_{o}^{2}}\over{8\pi G(1-\dot\phi^{2})}}
\label{eq:47}
\een
and the critical matter density becomes same as that of $\rho_{d0}$. 

Now from (\ref{eq:42}) with $K=0$, we have 
$({\dot a\over {a}})^2={C\over{a^{3}}}(1-\dot\phi^{2})$. We now take the negative 
square root of this equation so as to be consistent with observations. This will 
be borne out later. Therefore, 
\ben
{\dot a\over {a}}=-{{C^{1\over{2}}}\over{a^{3\over{2}}}}{(1-\dot\phi^{2})^{1\over{2}}}
\label{eq:48}
\een
where $C=\frac{8\pi GA}{3}$. Using equations (\ref{eq:20}) and (\ref{eq:48}) we get 
\ben
a(t)={C^{1\over{3}}}{({\dot\phi\over{\ddot\phi}})^{2\over{3}}}{(1-\dot\phi^{2})}\nonumber\\
\label{eq:49}
\een
Therefore the deceleration parameter for non-relativistic case with $K=0$
\ben
q(t)^{NR}=-\frac{a\ddot{a}}{\dot{a^{2}}}=\frac{Numerator}{Denominator}.
\label{eq:50}
\een
where,

\ben
Numerator=(1-\dot{\phi^{2}})[(1+20\dot{\phi^{2}})(\ddot\phi)^{4}\nonumber\\
+\dot\phi\phi^{(3)}(\ddot\phi)^{2}(1-4\dot\phi^{2})\nonumber\\
+5\dot{\phi^{2}}(\phi^{(3)})^{2}(-1+\dot{\phi^{2}})-3\dot{\phi^{2}}
\ddot{\phi}{\phi^{(4)}}(-1+\dot{\phi^{2}})]\nonumber\\
\label{eq:51}
\een
and
\ben
Denominator \nonumber\\
=2[(1-4\dot\phi^{2})(\ddot\phi)^{2}+\dot\phi{\phi^{(3)}}(-1+\dot\phi^{2})]^{2}\nonumber\\
\label{eq:52}
\een
We shall take $\phi^{(3)},\phi^{(4)}$ to be zero  also neglecting higher order
of $\dot\phi^{2}$, where $\phi^{(3)}$ is 
3rd order derivative with respect to time. Then the deceleration parameter for non-relativistic case becomes,
\ben
q(t)^{NR}=\frac{1}{2}(1+27\dot\phi^{2}+...)
\label{eq:53}
\een
Note that the choice of the sign of the square root (that leads to (\ref{eq:48}))
ensures that the  value of the deceleration parameter for the matter dominated 
era is as in standard cosmology i.e. when dark energy is absent. This value is 
$\frac{1}{2}$. Moreover, it can be checked that choice of a positive square root 
(leading to (\ref{eq:48})) will give  an imaginary scale factor which is unacceptable.
 
\subsection{Relativistic case (Radiation dominated Universe)}
For this case $p_{d}=\frac{\rho_{d}}{3}$ from (\ref{eq:40}) $\rho_{d}=\frac{B}{a^{4}}$ 
then modified Friedmann equations 
(\ref{eq:35}) and (\ref{eq:36}) becomes
\ben
{H^{2}\over{(1-\dot\phi^{2})}}+{K\over{a^{2}}}={8\pi G\over{3}}\frac{B}{a^{4}}
\label{eq:54}
\een
and 
\ben
2{\ddot a\over{a}}{1\over{(1-\dot\phi^{2})}}
+{H^{2}\over{(1-\dot\phi^{2})}}+{K\over{a^{2}}}
+2H{\dot\phi\ddot\phi\over{(1-\dot\phi^{2})^{2}}}\nonumber\\=-{8\pi G\over{3}}\frac{B}{a^{4}}
\label{eq:55}
\een
Considering the $K=0$ model of the Universe , the modified Friedmann equation (\ref{eq:54}) becomes 
$({\dot a\over a})^2={D\over a^4}(1-\dot\phi^{2})$. Again we take the negative square root 
of this equations from physical considerations to get  
\ben
{\dot a\over a}= -{D^{1\over 2}\over a^2}{{(1-\dot\phi^{2})^{1\over{2}}}}
\label{eq:56}
\een
where $D=\frac{8\pi GB}{3}$.
Again combining equations (\ref{eq:20}) and (\ref{eq:56}) we obtain
\ben
a(t)= D^{1\over{4}} {({\dot\phi\over{\ddot\phi}})^{1\over 2}}{(1-\dot\phi^{2})^{3\over 4}}
\label{eq:57}
\een
Therefore the deceleration parameter for relativistic case with $K=0$ is:
\ben
q(t)^{R}=-\frac{a\ddot{a}}{\dot{a^{2}}}=\frac{Numerator}{Denominator}
\label{eq:58}
\een
where,
\ben
Numerator=[(1+10\dot{\phi^{2}}-8\dot{\phi^{4}})(\ddot{\phi)^{4}}\nonumber\\
-3\dot{\phi^{2}}(\phi^{(3)})^{2}(-1+\dot{\phi^{2}})^{2}+2\dot{\phi^{2}}
(-1+\dot{\phi^{2}})^{2}\ddot\phi{\phi^{(4)}}]\nonumber\\
\label{eq:59}
\een
and 
\ben
Denominator=[(1-4\dot\phi^{2})(\ddot\phi)^{2}+\dot\phi{\phi^{(3)}}(-1+\dot\phi^{2})]^{2}
\nonumber\\
\label{eq:60}
\een
Finaly the deceleration parameter (neglecting as above higer order of $\dot\phi^{2}$ and 
higher order derivatives) for relativistic case is
\ben
q(t)^{R}=1+18\dot\phi^{2}+...
\label{eq:61}
\een 
Again, the choice of the sign of the square root  (leading to (\ref{eq:56})) 
ensures that the standard cosmology result is obtained for the deceleration 
parameter. This value is $1$. Also choice of a positive square root is ruled 
out to ensure reality of the scale factor.

\subsection{Dark energy dominated Universe}
For this case $p_{d}\simeq-\rho_{d}$ and from (\ref{eq:41}) $\rho_{d}=constant=W$, 
then the modified Friedmann equations (\ref{eq:35}) and (\ref{eq:36}) 
becomes
\ben
{\dot a^{2}\over{a^{2}}}{1\over{(1-\dot\phi^{2})}}+{K\over{a^{2}}}={8\pi G\over{3}}W
\label{eq:62}
\een
and
\ben
2{\ddot a\over{a}}{1\over{(1-\dot\phi^{2})}}
+{\dot a^{2}\over{a^{2}}}{1\over{(1-\dot\phi^{2})}}+{K\over{a^{2}}}
+2{\dot a\over{a}}{\dot\phi\ddot\phi\over{(1-\dot\phi^{2})^{2}}}\nonumber\\
=8\pi GW\nonumber\\
\label{eq:63}
\een
Again we consider $K=0$ model of the Universe, The modified Friedmann equation (\ref{eq:62})
 becomes
\ben
{\dot a\over a}=\alpha^{1\over {2}}{{(1-\dot\phi^{2})^{1\over{2}}}}
\label{eq:64}
\een
where $\alpha={8\pi GW\over{3}}=constant$.
Now combining equations (\ref{eq:20}) and (\ref{eq:64}) we obtain;
\ben
\frac{\ddot\phi}{\dot\phi}= -\alpha^{1\over {2}}{{(1-\dot\phi^{2})^{3\over{2}}}}.
\label{eq:65}
\een
Now from equation (\ref{eq:64}) we get the scale factor
\ben
a(t)=e^{\sqrt{\alpha}\int\sqrt{1-\dot\phi^{2}}dt}
\label{eq:66}
\een
Using above equation (\ref{eq:66}) we have the deceleration parameter for this case
\ben
q(t)^{dark}=-1+\frac{\dot\phi\ddot\phi}{\sqrt{\alpha}(1-\dot\phi^{2})^{3/2}}
\label{eq:67}
\een
Further, using (\ref{eq:65}) the deceleration parameter becomes
\ben
q(t)^{dark}=-1-\dot\phi ^{2}
\label{eq:68}
\een
As $\dot\phi ^{2}$ is positive the deceleration parameter is always negative.

Consider now an interesting situation. The dark energy density $\dot\phi ^{2} < 1$ 
for reasons mentioned before. Then expanding the binomial in (\ref{eq:65}) and keeping terms 
upto $O(\dot\phi^{2})$ , an approximate solution for the dark energy density is obtained as 
\ben
\dot\phi ^{2} = \frac{2\sqrt{\alpha}}{3\sqrt{\alpha} + 2 e ^{2\sqrt{\alpha}t}}
\label{eq:69}
\een

\vspace{1in}
\begin{figure}[h]
\centering
\includegraphics[scale=0.3]{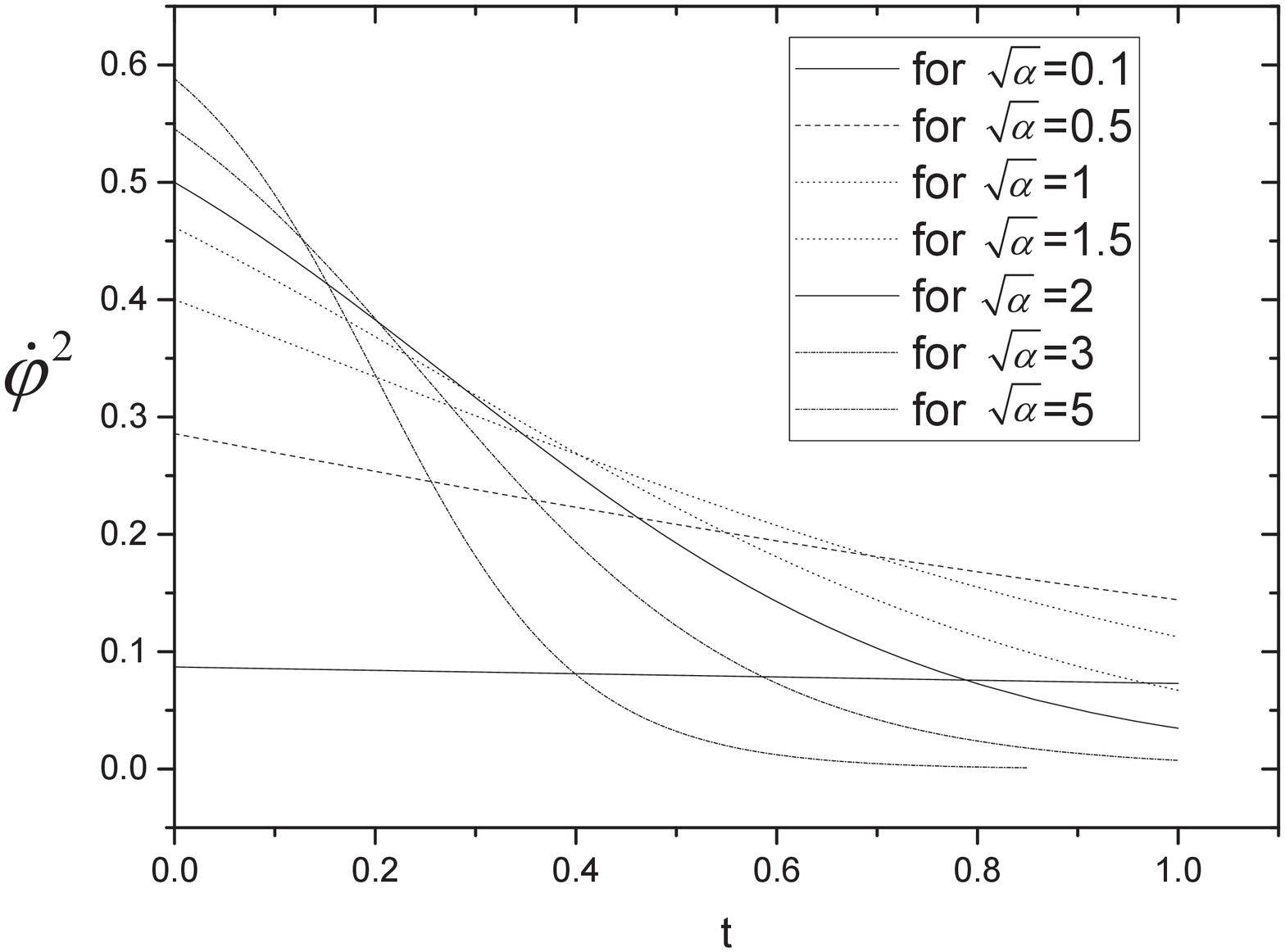} 
\caption{Variation of dark energy density with time for positive $\sqrt{\a}$
where values of $\sqrt{\a}$ are shown $down\rightarrow up$.}
\end{figure}

\begin{figure}[h]
\centering
\includegraphics[scale=0.3]{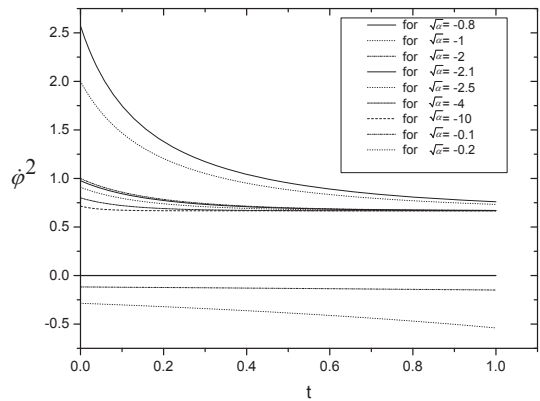} 
\caption{Variation of dark energy density with time for negative $\sqrt{\a}$ 
where values of $\sqrt{\a}$ are shown $up\rightarrow down$.}
\end{figure}
\vspace{1in}

Now  $\dot\phi ^{2} < 1$  applied  to  (\ref{eq:69}) means 
\ben
\sqrt{\alpha} + 2e^{2{\sqrt\alpha}t} > 0
\label{eq:70}
\een 
Note that if $\sqrt\alpha $ is  positive then (\ref{eq:70}) is  always satisfied for  all values of $t$. 
However, figure 1 shows that there is absolutely no agreement of the predicted values of dark energy density with the observed data \cite{planck1, planck2} at present epoch. So we reject this choice.

On the other hand, taking the negative square root for $\alpha$ gives encouraging agreement of predicted values for the dark energy density $\dot\phi^2$ with the observed value at present epoch {\textit viz.}, $0.6817$ \cite{planck1, planck2}. This is evident in figure 2. So we choose the negative square root. The best agreements  are obtained  for $-10\leq\alpha\leq -2.1$.

\section{Conclusion} 
In this work we have investigated the cosmological consequences of incorporating dark energy in an emergent gravity scenario. First we obtained the  analogues of the Friedman equations where the background metric is taken to be FLRW. 
Assuming the usual perfect fluid model for the universe, we next determined the total energy density. Finally, the cosmological implications were determined corresponding to various values of this energy . Our findings are as follows :   

(a)For total energy density greater than the pressure  the deceleration parameter $q(t)\approx\frac {1}{2} [1 + 27 \dot\phi ^{2}+...] > \frac{1}{2}$.   
(b)For total energy density equal to 3 times the pressure,  $q(t)\approx 1 + 18\dot\phi ^{2} +... >  1$  and
(c) for total energy density equal to the negative of the pressure (dark energy scenario), the deceleration parameter $q(t)< -1$. 

Note that for dark energy density $\dot\phi ^2=0$, the conventional results are retrieved. Our results indicate that many aspects 
of standard cosmology can be accommodated with the presence of dark energy  right 
from the beginning of the universe where the time parameter $t\equiv \frac{t}{t_{0}}$, $t_{0}$
being the present epoch. 


\begin{thebibliography}{99}
\bibitem{gor1}
V.Gorini,A.Kamenschik and U.Moschella, Phys.Rev. {\bf D67} 063509 (2003).
\bibitem{gor2}
V.Gorini,A.Kamenschik and U.Moschella and V.Pasquier ,arXiv:gr-qc/0403062 (2004).
\bibitem{riz} 
L.Rizzi,S.Cacciatori,V.Gorini,A.Kamenschik and O.F.Piatella, Phys.Rev {\bf D82} 027301 (2010).
\bibitem{kamen}
A.Y.Kamenschik,A.Tronconi and G.Venturi, Phys.Lett. {\bf B702} 191 (2011).
\bibitem{scherrer}
R.J. Scherrer, Phys.Rev.Lett.{\bf 93} 011301 (2004).
\bibitem{dg1}
D.Gangopadhyay and S. Mukherjee, Phys. Lett.{\bf B665} 121 (2008).
\bibitem{dg2}
D.Gangopadhyay, Gravitation and Cosmology {\bf 16} 231 (2010).
\bibitem{dg3}
D.Gangopadhyay and Goutam Manna, Euro.Phys.Lett. {\bf 100} 49001 (2012).
\bibitem{dg4}
Goutam Manna and D. Gangopadhyay, Eur. Phys. J. C {\bf 74} 2811 (2014).
\bibitem{armen1}
C.Armendariz-Picon, T.Damour and V.Mukhanov, Phys.Lett.{\bf B458} 209 (1999).
\bibitem{armen2}
C.Armendariz-Picon, V.Mukhanov and P.J.Steinhardt, Phys.Rev.{\bf D63} 103510 (2001).
\bibitem{armen3}
C.Armendariz-Picon and E.A.Lim, JCAP {\bf 0508} 007 (2005).
\bibitem{chiba}
T.Chiba, T.Okabe and M.Yamaguchi, Phys.Rev.{\bf D62} 023511 (2000).
\bibitem{cald}
R.R.Caldwell, Phys.Lett.{\bf B545} 23 (2002).
\bibitem{callan}
J.Callan, G.Curtis and J.M.Maldacena, Nucl.Phys.{\bf B513} 198 (1998).
\bibitem{rendall}
A.D.Rendall, Class.Quant.Grav.{\bf 23} 1557 (2006).
\bibitem{gibbons}
G.W.Gibbons, Rev.Mex.Fis.{\bf 49S1} 19 (2003);~~

\bibitem{wald}
R.M.Wald, in {\it General Relativity,The Univ.Chicago Press, (1984)}.
\bibitem{visser}
M.Visser,C.Barcelo and S.Liberati, Gen.Rel.Grav. {\bf 34} 1719 (2002);
\bibitem{babi1}
E.Babichev, V.Mukhanov and A.Vikman, JHEP {\bf 09}, 061 (2006).
\bibitem{babi2}
E.Babichev,M.Mukhanov and A.Vikman, JHEP {\bf 0802} 101 (2008).
\bibitem{babi3}
E.Babichev,M.Mukhanov and A.Vikman, WSPC-Proceedings, February 1, 2008.

\bibitem{vikman}
Alexander Vikman, {\it K-essence: Cosmology, causality and Emergent Geometry}, Dissertation an der Fakultat fur Physik,
Arnold Sommerfeld Center for Theoretical Physics, der Ludwig-Maximilians-Universitat Munchen, Munchen, den 29.08.2007.

\bibitem{born}
M.Born and L.Infeld,Proc.Roy.Soc.Lond {\bf A144}(1934) 425.

\bibitem{wein1}
S. Weinberg, {\it Gravitation and Cosmology}, Wiley Student Edition, John Wiley and Sons (Asia) Pte. Ltd., 2004.



\bibitem{wein2}
S. Weinberg, {\it Cosmology}, Oxford Univ. Press, 2008.
\bibitem{mukhanov}
V. Mukhanov, {\it Physical Foundations of Cosmology}, Cambridge University Press, 2005.

\bibitem{planck1}
Planck 2013 results. I. Overview of products and scientific results, Planck collaboration, arXiv.1303.5062.
\bibitem{planck2}
Planck 2013 results. XVI. Cosmological parameters,Planck collaboration, arXiv.1303.5076.











\end{thebibliography}
\end{document}